\documentclass[a4paper,11pt]{article}
\usepackage{pos}

\newcommand{\be}{\begin{eqnarray}}
\newcommand{\ee}{\end{eqnarray}}
\newcommand{\e}{e}
\newcommand{\ctwo}[2]{\left( \begin{array}{c} #1 \\ #2 \end{array} \right)}

\title{Tuning HMC parameters with gradients}

\author*[a]{James C. Osborn}
\affiliation[a]{Computational Science Division, Argonne National Laboratory,\\
  9700 S Cass Ave, Argonne, IL 60439, USA}
\emailAdd{osborn@anl.gov}

\abstract{We investigate the effectiveness of tuning HMC parameters using
information from the gradients of the HMC acceptance probability with
respect to the parameters. In particular, the optimization of the
trajectory length and parameters for higher order integrators will be
studied in the context of pure gauge and dynamical fermion actions.}

\FullConference{The 40th International Symposium on Lattice Field Theory (Lattice 2023)\\
July 31st - August 4th, 2023\\
Fermi National Accelerator Laboratory\\}

\begin{document}
\maketitle

Gauge field generation in lattice field theory (LFT) generally relies on
the Hybrid (or Hamiltonian) Monte Carlo (HMC) algorithm \cite{DUANE1987216}
which integrates the Hamiltonian equations of motion to generate
a new trial configuration.
Higher-order integration schemes have been shown to improve the
efficiency of HMC simulations for LFT applications \cite{Takaishi_2006}.
These higher-order schemes can have several parameters that, in general,
should be tuned for the specific application.
Furthermore, the action can have preconditioners, such as that by Hasenbusch
\cite{Hasenbusch_2001}, which introduce more free parameters to tune.
While the tuning can be, and often is, done by hand by monitoring the
acceptance rate and/or norms of the force terms,
this can become tedious as the number of parameters increases.
Here we investigate automating the tuning process using techniques
commonly employed in machine learning applications.

\section{HMC}

The Hybrid Monte Carlo method proceeds by first choosing a
random Gaussian momentum field, $p$, which is associated with the gauge field degrees
of freedom, $U$.
Then the equations of motion for the Hamiltonian
$H(p,U) = \frac{1}{2} p^2 + S(U)$,
with $S(U)$ the lattice action to be sampled from,
are approximately integrated for some integration time $\tau$.
This produces a new set of fields
\be
\ctwo{p'}{U'} = F_\alpha \ctwo{p}{U}
\ee
where $F_\alpha$ is the operator implementing the integration scheme and
$\alpha$ represents the integration time, $\tau$, and any other tunable integration
or action parameters.
The new configuration is conditionally accepted with the probability
\be
P_{a}(p', U', p, U) = \min[ 1, \exp(-\Delta H) ]
\ee
with
\be
\Delta H = H(p', U') - H(p, U) ~.
\ee

\subsection{Improved integrators}

For simplicity we'll consider integrators with a single force type
which could be either a gauge force or a combined gauge and fermion force.
The integrator $F_\alpha$ is an approximation of $\exp(\tau[A+B])$
where the operator $A$ is the gauge field update and
$B$ is the momentum update using the force term.
This update can be divided into $n$ copies of a base integrator as
$\exp(\tau[A+B]) = [\exp(\epsilon[A+B])]^n$ with $\epsilon = \tau/n$.

To start, we consider a simple two-step base integrator
given by the approximation
\be
\e^{\epsilon (A+B)} \approx
 \e^{\epsilon\lambda A}
 \e^{(\epsilon/2) B}
 \e^{\epsilon(1-2\lambda) A}
 \e^{(\epsilon/2) B}
 \e^{\epsilon\lambda A}
\ee
where $\lambda$ is a free parameter to tune.
Omelyan, et al. \cite{Omelyan_2002} gave an optimal value for this parameter of
$\lambda \approx 0.193$,
under the assumption that the two operators that appear in the leading order error
term have equal magnitudes.
While this assumption produces a reasonably good integrator, for LFT applications,
the magnitudes may not be very close, and one can typically reduce the error by
tuning the parameter for specific actions \cite{Takaishi_2006}.

One can also consider higher order integrators, such as force-gradient
variants.
The force-gradient integrator we will consider here is
\be
\e^{\epsilon (A+B)} \approx
 \e^{\epsilon\theta A}
 \e^{\epsilon\lambda B}
 \e^{(\epsilon/2)(1-2\theta) A}
 \e^{\epsilon(1-2\lambda) B + \epsilon^3\chi C}
 \e^{(\epsilon/2)(1-2\theta) A}
 \e^{\epsilon\lambda B}
 \e^{\epsilon\theta A}
\ee
where $C = [B,[A,B]]$ is the force-gradient term.
For simplicity we implement this term using the method introduced in \cite{Yin:2012rX}.
This force-gradient update has three parameters, $\theta$, $\lambda$, and $\chi$,
that can be tuned, however the cancellation of the order $\epsilon^3$ errors
places a constraint on two of the parameters, leaving only one free
parameter.
Here, for testing purposes, we will leave all three unconstrained when tuning
in order to evaluate the tuning process, and to allow for a compromise
in reducing the order $\epsilon^5$ error at the expense of not completely canceling
the order $\epsilon^3$ errors.

In full LFT simulations with dynamical quarks, there
can also be different integration scales used for the gauge force and the
fermion forces, using a recursive integrator \cite{Sexton_Weingarten_1992},
which can introduce many more parameters, although we will
not consider those here.
As the number of free parameters in the integration increases,
the prospect of automated tuning becomes a particularly attractive
alternative to the conventional methods of tuning by hand.

\section{Loss function}

To measure the efficiency of the integrator, we consider a cost function based
on the effective integration time.
For a set of $N$ HMC update steps, each of integration time $\tau$,
the total effective integration time is
\be
\tau \sqrt{\langle P_{a} \rangle N}
\ee
where $\langle P_{a} \rangle$ is the average acceptance probability
and the square root comes from the random walk due to choosing a new momentum
at the beginning of each HMC update.
The number of update steps needed to go $T$ effective integration time units is then
\be
N_T = \frac{T^2}{\langle P_{a} \rangle \tau^2} ~.
\ee
We can then define the cost of updates for an effective integration time $T$ as
the cost per update of length $\tau$, times $N_T$.
For simplicity we set the cost per update based on the number of force evaluations,
$N_{force}$, and take $T=1$.
This gives the final cost metric
\be
\label{cost}
Cost = N_T N_{force} = \frac{N_{force}}{\langle P_{a} \rangle \tau^2} ~.
\ee

We want to minimize the cost function (\ref{cost}), however this quantity has an average
in the denominator which makes it inconvenient to work with.
Instead we minimize the loss function
\be
\label{loss}
Loss = - \langle P_{a} \rangle \tau^2
\ee
which has a similar effect and is much easier to work with.
For the minimization we use machine learning (ML)
methods and use the gradient of the loss function
to update the HMC parameters.
We use the Adam optimizer \cite{adam} and accumulate the gradient during the update using
a single update stream (batch size of one).
Since the Adam optimizer already accumulates the gradient information during
the optimization process, we do not need to explicitly include the ensemble
average of $P_{a}$ in the loss function, and instead just calculate the gradient
of $P_a$ for the current HMC update.
We then update the tunable parameters based on the accumulated gradient
at the end of every HMC update.

\section{Calculating gradients}

Optimizing the parameters $\alpha$ requires calculating the gradient of the loss function
in Eq. (\ref{loss}) with respect to each of the parameters $\alpha$.
This is typically done through back propagation, which starts from the definition
of the loss function and evaluates the chain rule back through all the intermediate fields.

The main task is evaluating the gradient of $P_{a}$,
which is a function of $H(p',U') \equiv H'$.
Note that if $\Delta H \le 0$ then the gradient of $P_a$ is zero.
In the other case, we consider the integrator as a series of individual
update steps.
For example, for $n$ copies of the two-step integrator, written in operator notation,
we have
\be
\ctwo{p'}{U'}
= \left[ e^{\epsilon \lambda A} e^{(\epsilon/2) B} e^{\epsilon (1-2\lambda) A}
  e^{(\epsilon/2) B} e^{\epsilon \lambda A} \right]^n \ctwo{p}{U} ~.
\ee
We can calculate the gradients from chain rule
\be
\label{cr}
\frac{\partial H'}{\partial \alpha} =
\sum_k \frac{\partial H'}{\partial q_k} \left. \frac{q_k}{\partial \alpha}
\right|_{q_{k-1}}
\ee
where
\be
q_k = \left( \begin{array}{c} p_k \\ U_k \end{array} \right)
\ee
is the field state after the $k$th integrator step (either applying $A$ or $B$)
starting with $q_0 = (p, U)$.
The first term in the chain rule, Eq. (\ref{cr}),
is accumulated from the end of the update
working backwards using
\be
\frac{\partial H'}{\partial q_k} =
\frac{\partial H'}{\partial q_{k+1}} \frac{\partial q_{k+1}}{\partial q_{k}}
\ee
and the second term is the gradient of
the individual update step with respect to the parameters keeping the input
fields, $q_{k-1}$, fixed.
One then needs to calculate the gradients for the individual update steps which
we do not include here.
One issue with the calculation of the gradients is the need to save all the intermediate
fields during an update step.
This requires a large increase in the memory usage, however since the HMC is typically
scaled out to many nodes to reduce the time per update as much as possible,
memory use is not likely to be a limiting factor.

\section{Results}

The code was implemented using the QEX LFT framework \cite{qex,Jin:2017nN}.
We are using a simple ``tape'' implementation which saves a list of operations
forming the HMC update.
The tape is then run in the forwards direction to perform the HMC update,
and can be run in the backwards direction
to produce the gradient of the full HMC update step.
All runs were done on a relatively small $12^3 \times 24$ lattice
running on a single desktop.

\subsection{Pure gauge results}

The initial tests were performed on a pure gauge theory
with a plaquette action at $\beta = 5.6$.
For all runs we started from a thermalized configuration and then ran
200 tuning updates and 400 measurement updates.
The number of tuning updates was taken as a constant for all runs
for simplicity and was chosen to ensure that all runs were sufficiently
tuned.
In many cases the number of tuning updates could have been lowered
without impacting results.

\begin{figure}
  \centering
  \includegraphics[width=0.50\textwidth]{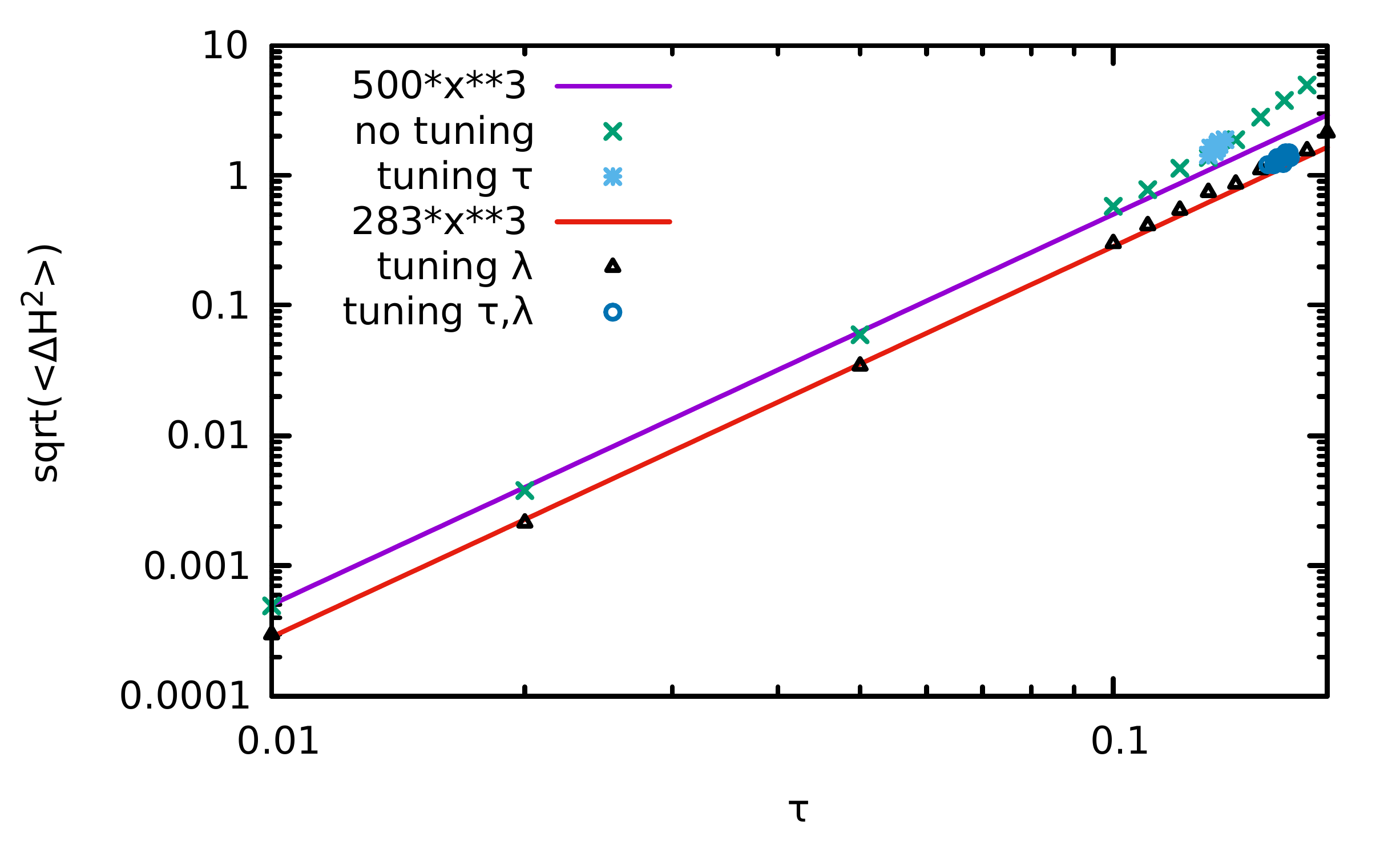}
  \includegraphics[width=0.49\textwidth]{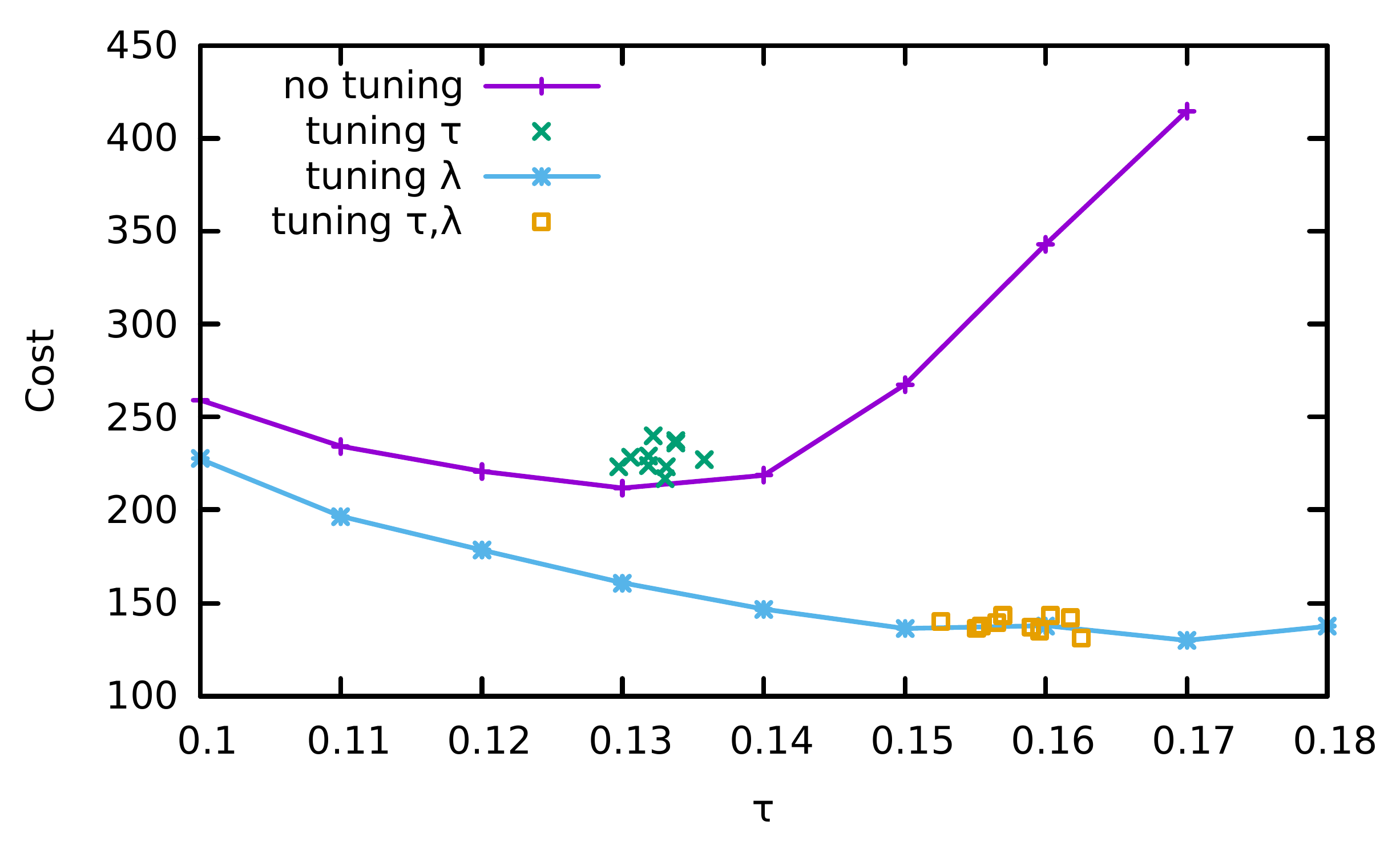}
  \caption{\label{dababa1}
    Pure gauge simulation with single copy ($n=1$, $\epsilon = \tau$) of a
    two-step integrator with and without tuning on the trajectory length,
    $\tau$, and the integrator parameter $\lambda$.
  Left: RMS integration error versus trajectory length.
  Right: Cost versus trajectory length.
  }
\end{figure}

We first examined tuning HMC using an integrator with a single copy of a two-step
update with the field update first.
Here there are two parameters we can tune, the trajectory length, $\tau$,
(which is the same as the step size, $\epsilon$, for a single copy)
and the integrator parameter $\lambda$.

On the left panel of Figure \ref{dababa1} we show results for the root mean squared (RMS)
integrator error,
$\sqrt{\langle\Delta H^2\rangle}$,
versus $\tau$ for tuned and untuned runs.
The green x's are for no tuning, using the Omelyan, et al. value of $\lambda$.
The error scales as $\tau^3$ for small $\tau$
as expected for a second order integrator, and is demonstrated by the purple line.
If we tune $\tau$, then we end up with the points marked with blue stars,
which are at larger values of $\tau$ where the asymptotic $\tau^3$
scaling starts to break down.
Each star represents an independent tuning run.
The black triangles are for tuning $\lambda$ at each value of $\tau$.
Again this scales as $\tau^3$ for small $\tau$, but now with a smaller coefficient
as shown by the red line.
The blue circles are the values obtained from runs tuning both $\lambda$ and $\tau$.
Again the tuned value of $\tau$ ends up right around the value where the asymptotic
scaling with $\tau$ starts to break down.

On the right panel of Figure \ref{dababa1} we show results for the cost function,
Eq. (\ref{cost}), versus $\tau$ for tuned and untuned runs.
For the case where $\lambda$ wasn't tuned, we can see that tuning $\tau$ was able to find
values of $\tau$ that were near the minimum of the cost function.
For the case of tuned $\lambda$, the minimum of the cost is very shallow, so the spread
of the runs with tuned $\tau$ is larger, but the values found were still close
to optimal.

\begin{figure}
  \centering
  \includegraphics[width=0.49\textwidth]{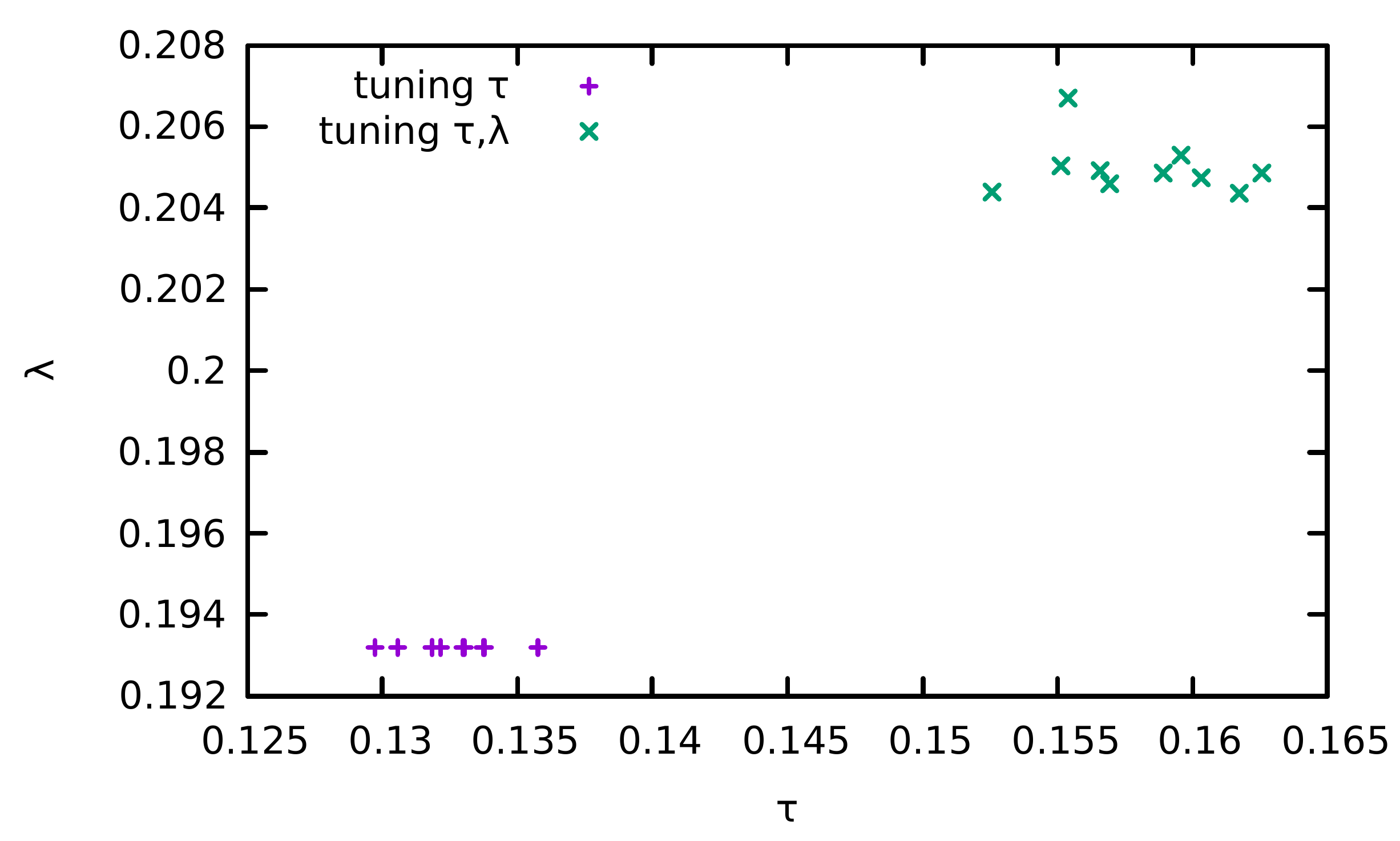}
  \includegraphics[width=0.49\textwidth]{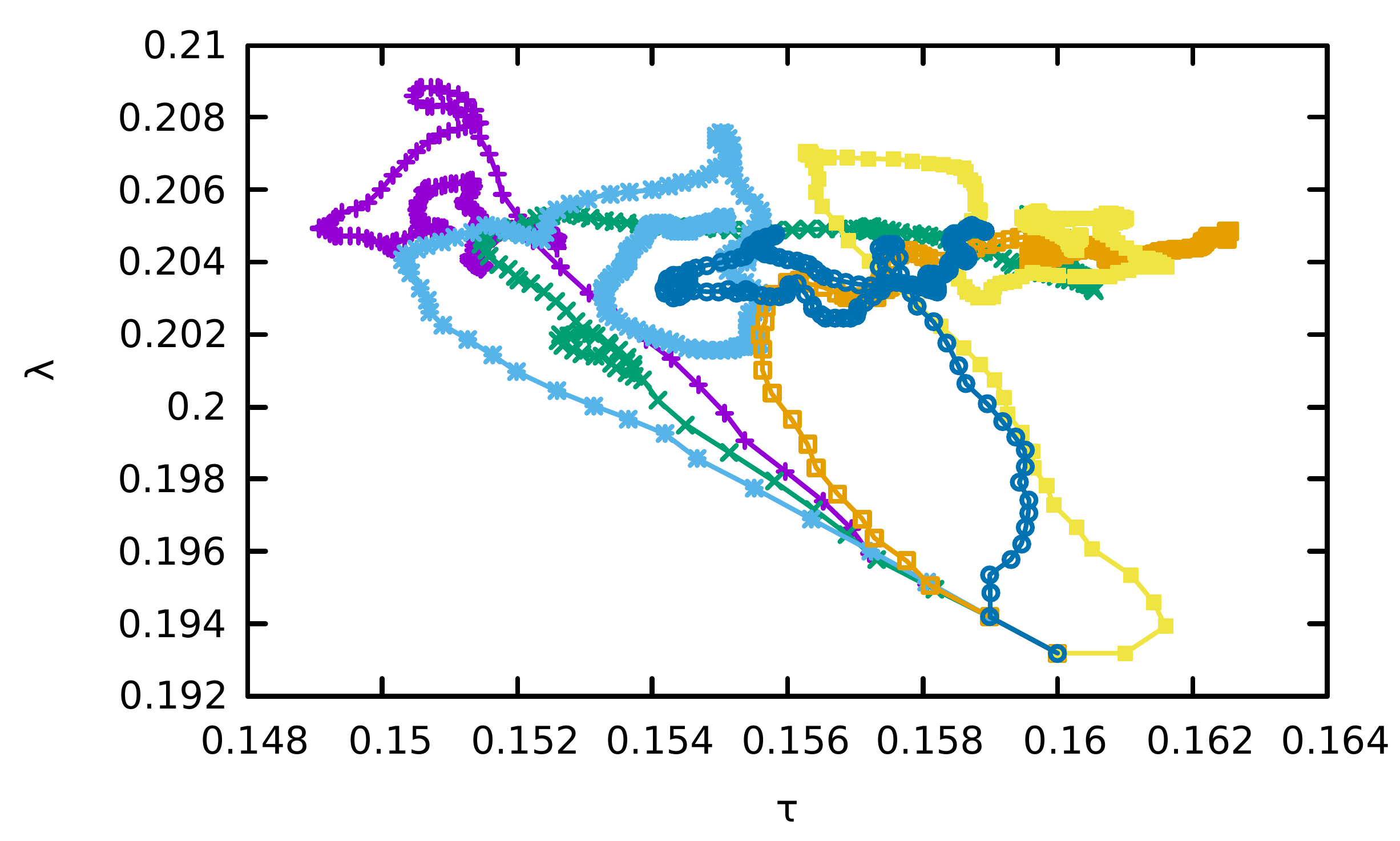}
  \caption{\label{pababa1}.
  Pure gauge simulation with single copy ($n=1$, $\epsilon = \tau$) of a two-step integrator.
  Left: Tuned parameters.
  Right: Parameter paths during tuning.
}
\end{figure}

In the left panel of Figure \ref{pababa1} we show the final values of the tuned
parameters for several tuning runs with and without tuning $\lambda$.
The tuned values are fairly consistent within each case, but there is a large
difference between the two cases.

In the right panel of Figure \ref{pababa1} we show the trajectories
taken in the parameter space during tuning when both parameters are
tuned, all starting from the same point.
While the trajectories may look very different, the final values are fairly consistent.
We also saw a similar convergence when using different starting parameters.

\begin{figure}
  \centering
  \includegraphics[width=0.49\textwidth]{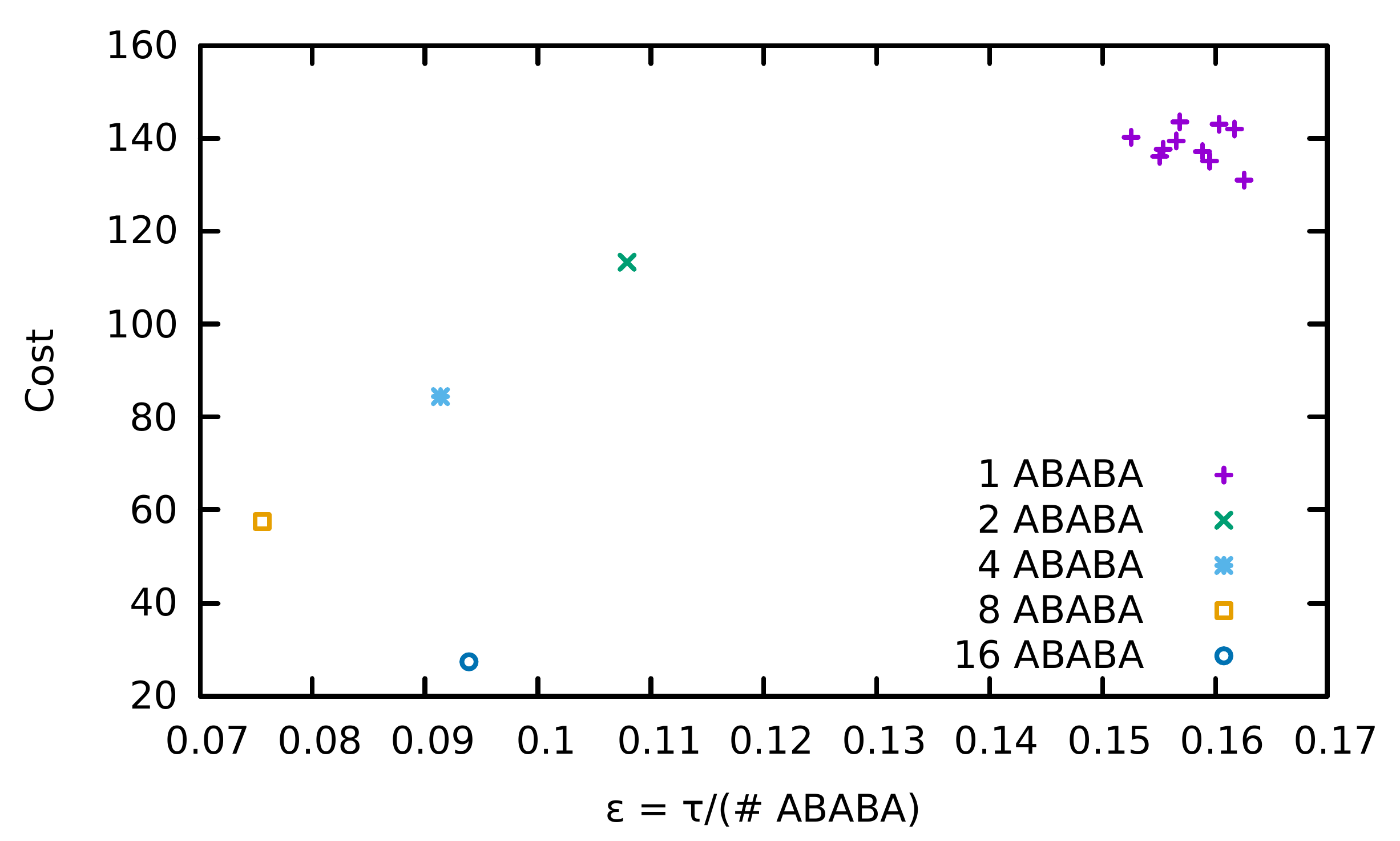}
  \caption{\label{cababan}
    Cost versus integrator step size for pure gauge simulation with multiple copies
    ($n =$ \# ABABA) of a two-step integrator with all parameters tuned.}
\end{figure}

In Figure \ref{cababan} we show the cost versus step size for the fully tuned HMC
with an integrator containing different numbers of copies of the two-step integrator.
We can see that increasing the number of copies decreases the cost up to the 16 copies
used here.
Also, the tuned value of $\epsilon$ varies with the number of copies.

\begin{figure}
  \centering
  \includegraphics[width=0.50\textwidth]{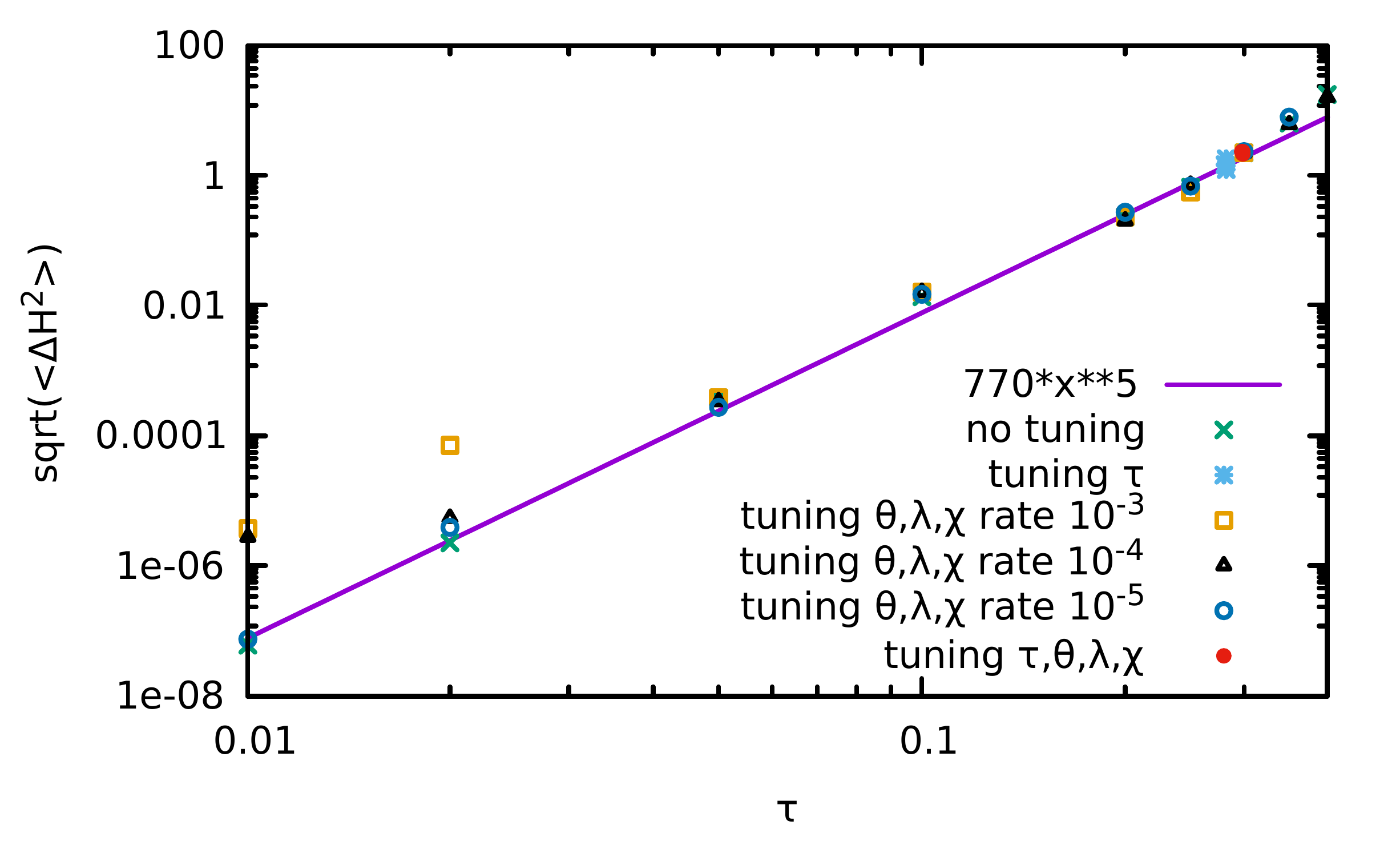}
  \includegraphics[width=0.49\textwidth]{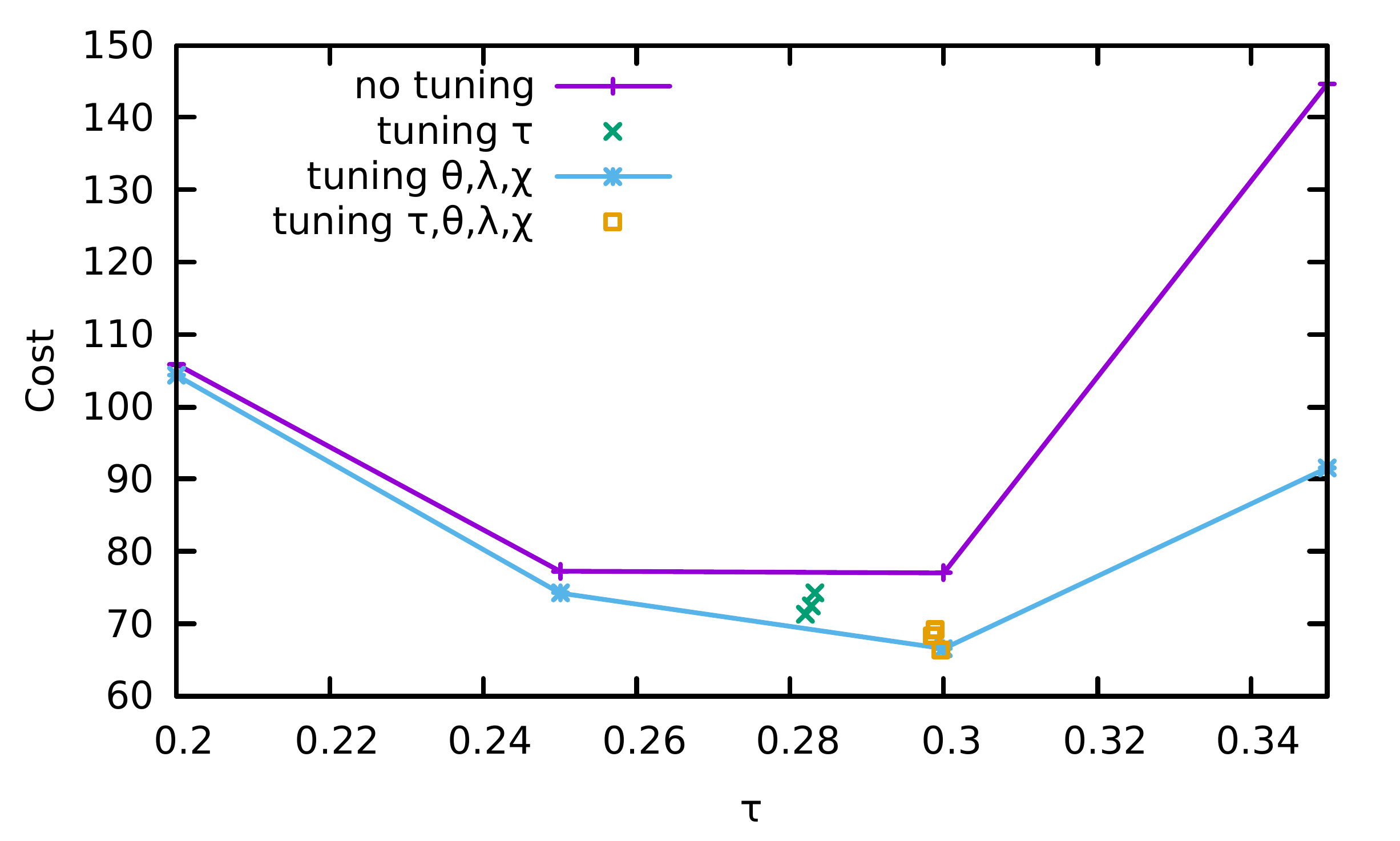}
  \caption{\label{dabacaba1}
    Pure gauge simulation with single copy ($n=1$, $\epsilon = \tau$) of a force-gradient
    integrator.
  Left: RMS integration error versus trajectory length.
  Right: Cost versus trajectory length.}
\end{figure}

In the left panel of Figure \ref{dabacaba1} we show results for the
RMS integration error versus $\tau$ for the pure gauge action using
the force-gradient integrator.  We found that the tuning rate used in
the Adam optimizer had a large effect on the final result when $\tau$
was held fixed at a small value.  We had to decrease the tuning rate
from the default of $10^{-3}$
to $10^{-5}$ in order to see the expected $\tau^5$ scaling of the
error when tuning the parameters.  However the error with the untuned
parameters, did achieve the expected scaling and performed at least as
well as the tuned case.  So in this case, tuning the force-gradient
integrator did not seem to be necessary.

In the right panel of Figure \ref{dabacaba1} we show results for the
cost versus $\tau$ for the pure gauge action using the force-gradient
integrator.  The runs with tuning $\tau$ were consistently able to
find the minimal cost value as seen from comparing to the runs with
different fixed values of $\tau$.

\subsection{Staggered quark results}

Next we performed tests using a single copy (4 tastes) of unimproved staggered quarks
with the plaquette gauge action.
The lattice volume, $12^3 \times 24$, and $\beta = 5.6$ were the same as in
the pure gauge runs.
The staggered mass was set to $m = 0.04$ which gives a pion mass
of about $0.5$ in lattice units.
Since the fermion force is much more expensive than the gauge force, we only count
the number of fermion force computations in calculating the cost function.
We also put the gauge and fermion forces on the same integrator timescale for
simplicity, instead of using a nested integrator.

\begin{figure}
  \centering
  \includegraphics[width=0.50\textwidth]{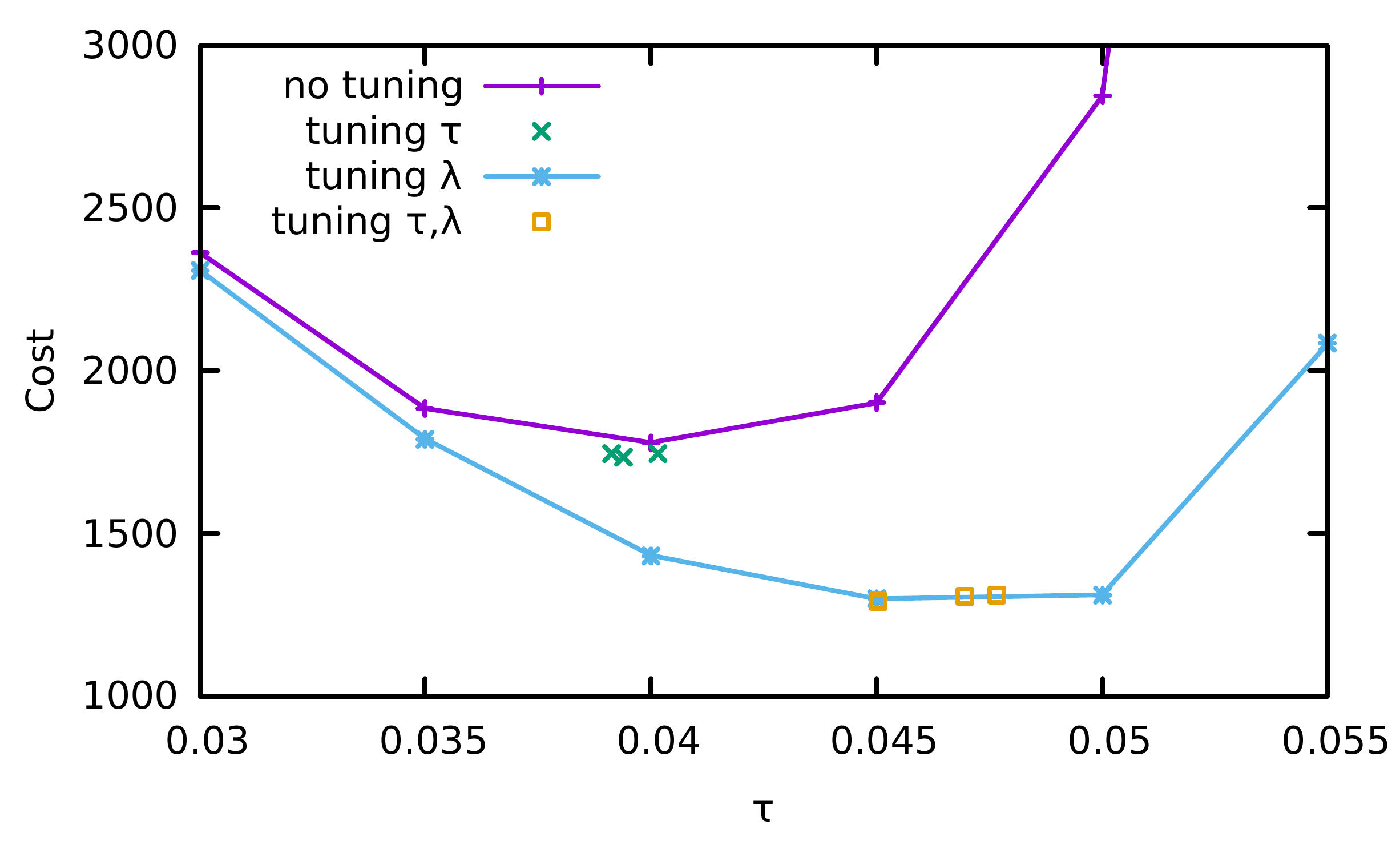}
  \includegraphics[width=0.49\textwidth]{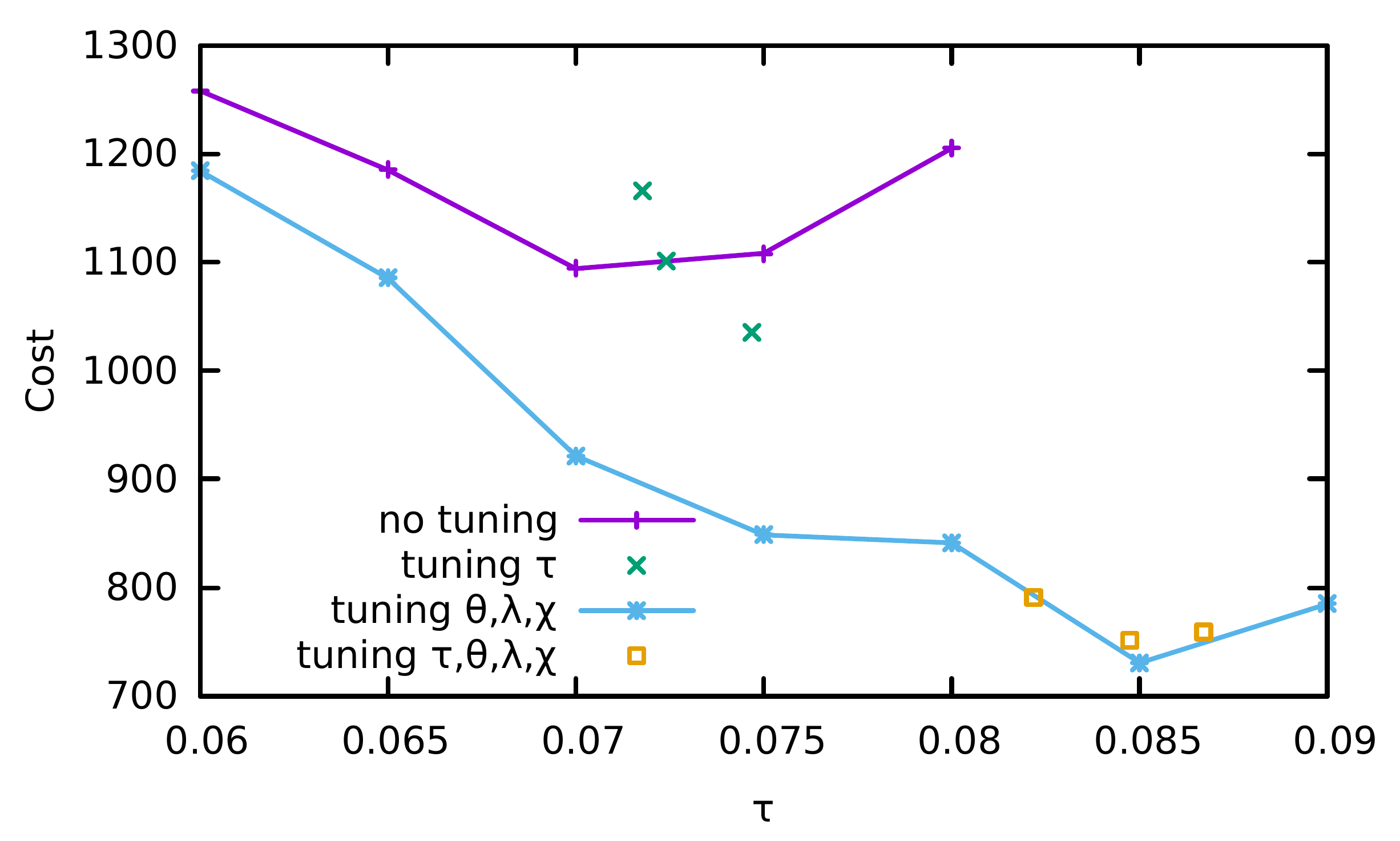}
  \caption{\label{stagcababa1}
    Cost versus trajectory length for staggered quark simulations with a
    single integrator copy ($n=1$, $\epsilon = \tau$).
  Left: Two-step integrator.
  Right: Force-gradient integrator.}
\end{figure}

Figure \ref{stagcababa1} shows the cost versus $\tau$ for staggered
fermion HMC with the two-step (left panel) and force-gradient (right panel) integrators.
Note that the force-gradient integrator can be generalized such that the integrator
parameters $\lambda$ and $\chi$ have different values for the gauge and fermion forces,
and we allowed them to be tuned independently.
We find that tuning $\tau$ is able to reliably find the minimal cost and that
tuning the other HMC parameters gives a noticeable improvement.
Note that for the case of staggered quarks, as opposed to pure gauge,
we see a significant improvement when tuning the force-gradient integrator.

\begin{figure}
  \centering
  \includegraphics[width=0.50\textwidth]{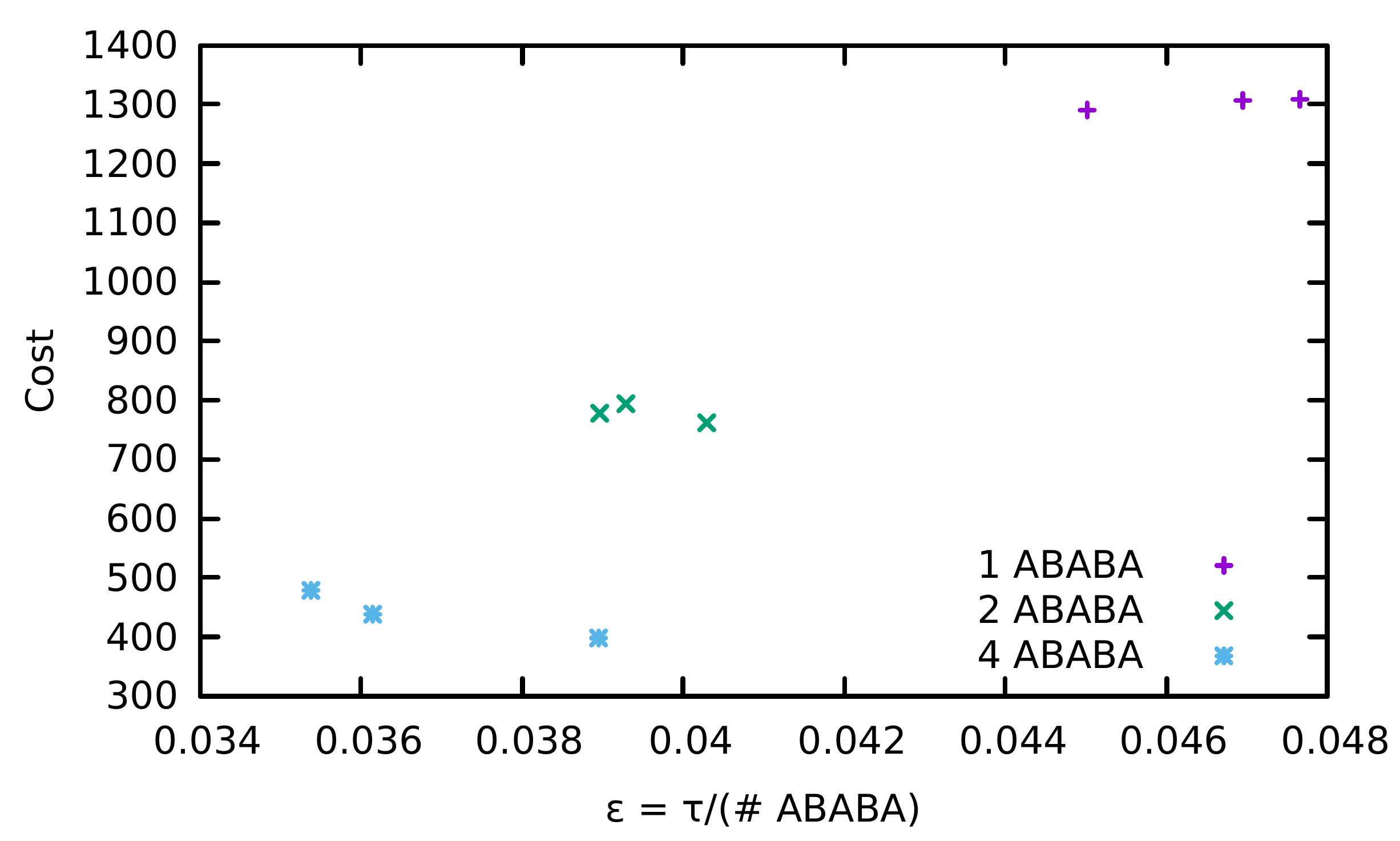}
  \caption{\label{stagcababan}
    Cost versus integrator step size for staggered quark simulations
    with multiple copies
    ($n =$ \# ABABA) of a two-step integrator with all parameters tuned.}
\end{figure}

In Figure \ref{stagcababan} we show the cost versus step size for staggered HMC
with multiple copies of the two-step integrator.
As with the pure gauge case, we see that the cost decreases as the number
of copies increases up to the 4 copies tested.
Note that for two copies of the two-step integrator, the cost is similar
to that of the tuned force-gradient integrator (which has the same
number of fermion force evaluations), with the force-gradient
cost function being slightly lower.
Without tuning however, the force-gradient integrator is much worse than two
copies of the two-step integrator.
As the volume increases we expect that the force-gradient integrator will perform
even better, but only if properly tuned.

\section{Summary}

We explored implementing and optimizing HMC simulations using gradient information
with methods borrowed from ML applications.
Overall we found that tuning HMC parameters using gradient information works well,
at least for the cases tested here.
In most cases tuning the HMC parameters made the HMC more efficient
with the only exception being the case of pure gauge action with
the force-gradient integrator.
Using the gradients is a very convenient way to tune HMC, once the initial
investment to develop the implementation has been done.

We also found that the force-gradient integrators may work much better when
tuned, at least in the case of staggered fermions, and
could be competitive with other integrators even on small volumes.
We plan to continue testing this approach with other actions (improved gauge,
improved fermions and with rooting)
and with tuning the mass parameters in Hasenbusch preconditioners.

This research was supported by the Exascale Computing Project
(17-SC-20-SC), a collaborative effort of the U.S. Department of Energy
Office of Science and the National Nuclear Security Administration.
This work was performed at Argonne National Laboratory which is
supported under Contract DE-AC02-06CH11357.

\bibliography{refs}
\bibliographystyle{plain}

\end{document}